\documentclass[11pt]{article}

\usepackage{graphicx}

\title{ Partial Decoherence and Thermalization through Time-Domain Ergodicity}

\author{Robert Englman$^a$ and Asher Yahalom$^{b,c}$  \\
$^a$Soreq NRC, Yavne 81800, Israel\\
$^b$ Isaac Newton Institute for Mathematical Sciences,\\
 20 Clarkson Road, Cambridge CB3 0EH, United Kingdom\\
$^c$Ariel University, Ariel 40700, Israel\\}
\begin{document}

\newcommand{\beq} {\begin{equation}}
\newcommand{\enq} {\end{equation}}
\newcommand{\ber} {\begin {eqnarray}}
\newcommand{\enr} {\end {eqnarray}}
\newcommand{\eq} {equation}
\newcommand{\eqs} {equations }
\newcommand{\mn}  {{\mu \nu}}
\newcommand{\sn}  {{\sigma \nu}}
\newcommand{\rhm}  {{\rho \mu}}
\newcommand{\sr}  {{\sigma \rho}}
\newcommand{\bh}  {{\bar h}}
\newcommand {\er}[1] {equation (\ref{#1}) }
\newcommand {\ern}[1] {equation (\ref{#1})}
\newcommand{\mbf} {{ }}
\newcommand {\Del} {\Delta}
\newcommand {\SE} {Schr\"{o}dinger equation}

 \maketitle

 \begin {abstract}
An approach, differing from two commonly used methods  (the
stochastic \SE \ and the master equation \cite
{Schlosshauer,BieleA}) but entrenched in the traditional density
matrix formalism,
 is developed in a semi-classical setting, so as to go from the solutions  of the time dependent
 \SE~ to decohering and thermalized states. This is achieved by utilizing the time-ergodicity,
rather than the sampling- (or ensemble-) ergodicity, of physical systems.

 We introduce the formalism through a study of the Rabi model (a two level system coupled to an oscillator)
 and show that our semi-classical version exhibits, both qualitatively and quantitatively, many features
  of state truncation and  equilibration \cite {AllahverdyanBN}. We then study the time evolution of
  two  qubits in interaction with a bosonic environment,
such that the energy scale of one qubit is much larger, and that of the other much smaller than
the environment's energy scale. The small energy qubit decoheres to a mixture, while the high energy
qubit is protected through the adiabatic theorem. However, an inter-qubit coupling generates
an overall decoherence and leads for some values of the coupling to long term revivals in the state occupations.

\end{abstract}
\section{Introduction}
Safe and reliable manipulation of quantum states (as is envisaged
in a quantum computer) depends on the possibility of error-free
and stable quantum systems when left alone, except for the
inevitable interaction with the environment. The devil is in
decoherence and numerous works have been devoted to estimate,
minimize, circumvent it or to correct for it \cite {Schlosshauer}.
Distinct from the direct approaches to provide somehow remedies
for decoherence , several avenues have been explored in which the
quantum system maintains coherence due to the Hamiltonian the
defines it. One of the earliest works related to the subject is by
Kubo \cite {Kubo}, in which hints for the approach taken in the
present paper can be found. While decoherence and dissipation are
terms very close to each other, "dissipationless decoherence" was
also considered \cite{GangopadhyayKD} and decoherence-free
subspaces in the Hilbert- space were studied in \cite {LidarW}. In
more recent works such subspaces were identified, manifesting
"partial decoherence", through symmetry-based discrimination
between parts of the Hilbert-space \cite
{AharonyED}-\cite{AharonyGTED}.

The present work also treats partial decoherence, but differs
 from the previous in that, rather than throwing the burden of
 discrimination on a specially contrived Hamiltonian, it finds
 discrimination between Hilbert subspaces more generically, through their having different energy
 scales. A simple physical example of this is an atomic system in the presence
 of a magnetic field of 1 tesla, for which the electronic
 spins separate to about 10 cm$^{-1}$  and the nuclear spins to
 about 10$^{-2}$ cm$^{-1}$.
 As already indicated, an allied idea was briefly noted
 by Kubo \cite{Kubo}, who differentiated between the cases of fast
 and slowly modulated frequencies of the relaxing system. A
 further idea borrowed in  the present work from that paper (and
 indeed from other treatments involving "ergodicities") is
 equating ensemble averages  with long time averages \cite {Farquhat}.

 The proposed semi-classical formalism (which is the main novelty  of this work)
 is introduced and tested in section 2 on the single qubit(or $\frac{1}{2}$ spin)-single boson (Rabi)
 model \cite{Rabi}. This was
  thoroughly treated algebraically \cite {IrishGMS}-\cite{Ziegler} and applicatively: for single trapped ions
  \cite{LeibfriedBMW}, chiral molecules in a three-level system \cite {ThanopulosPK}, Josephson junctions
  \cite {SornborgerCG}, a single photon coupled to to a superconducting (SC) qubit \cite{WallraffEA},
  the Bloch-Siegert shift in a SC flux qubit\cite{FornDiazEA}; all these  somewhat with a long term view
  of decoherence-ridden quantum computing.
The case of two qubits, in interaction with a single classical oscillator and having largely differing Zeeman
splitting energies,
 is considered  in section 3 and is the essential motivation for this work. The two qubit case featured in \cite
 {SteffenEA} and was recently treated
  algebraically in \cite {YangZZ} in cases amenable to adiabatic treatment,
 but when the two qubits have identical splitting energies.

\section {Decoherence in a Semi-classical Rabi model}

Here the spin-vibration Hamiltonian \beq H(t,a)=
e\sigma_z+k\sigma_x \sin(\omega t+\alpha_{a}) ~~~(\omega \to
1)\label{H0}\enq describes our two-parameter system, involving a
Zeeman-split 1/2-spin [represented by the Pauli matrices
  $ \sigma_z=  \left( \begin{array}{cc}
  1 &  0 \\
   0 & -1
  \end{array} \right)$, $ \sigma_x=  \left( \begin{array}{cc}
  0 &  1 \\
   1 & 0
  \end{array} \right)$]
and a classical vibrator, whose frequency $\omega$ is equated to
1, thus setting the time (t) scale and the
 energy scales of the splitting $2e$ and of the spin-vibration coupling $k$.
  $\alpha_{a}$ is an initial phase of the classical vibrator, whose value is
  specified by the indexing parameter $a$.
    The Hamiltonian of the classical vibration is not needed.
In our procedure the time dependent Schrodinger equation (TDSE)  [$i\frac{d \psi_{a}(t)}{dt}=H(t,a)\psi_{a}(t)]
$\footnote{We use units in which $\hbar=1$} is solved numerically
with some fixed initial conditions.

\subsection{Programmatic summary of the three steps to construct the density matrix $\rho_{mn}$}

\begin{enumerate}

\item We adopt the von Neumann definition \cite{Neumann}-\cite{EnglmanY2004var}:
 \beq
 \rho_{mn}(t) = \frac{1}{(\sum _a 1)} \sum_{a} <m|\psi_{a}(t)><\psi_{a}(t)|n>
 \label{rho0}
 \enq In this definition the summation index $a$ represents the values of all coordinates, variables etc. external to the
 system (e.g., those of the environment affecting the system) and appearing also in the Hamiltonian.
 Thus the set ${\psi_{a}(t)}$ for all $a$'s forms a time dependent {\it ensemble} of states. The degrees
  of freedom of the system themselves are implicit (not written out) in $\psi_a (t)$.

  \item We solve only for a single external condition  thus dispensing with the $a$ index in the wave function,
 but obtain $\rho(t)$ as the average over an adequate set of adjacent times:
 \beq
 \rho_{mn}(t)= \frac{1}{2 \Delta t}\int_{t-\Delta t}^{t+\Delta t}d\tau <m|\psi(\tau)>
 <\psi(\tau)|n>\label{rho2}
 \enq
  This should be equivalent to \er{rho0} if the ergodic hypothesis holds for the duration $2\Delta t$.
  (It also represents a considerable simplification in numerics, since the TDSE is only solved once,
  specifically for $\alpha=0$.
 To justify the replacement of step 1 by step 2 we note that in all cases considered, numerically computed non diagonal density matrix elements were several orders of
  magnitude smaller than the diagonal ones. Thus decoherence,  which is the "truncation" of \cite {AllahverdyanBN},
   was achieved. The time-averaging method also avoids the notorious "initial slippage" problem \cite{GaspardN}.
   More detailed motivation for step 2 is given in the Discussion section, after the reader has become informed of the
   proposed method).

 \item For the basis  $n,m$ set we have chosen two alternative representations: (a) the spin
 eigenstates,
 up: $\left( \begin{array}{c}
  1  \\
  0   \end{array} \right)$ and down:$\left( \begin{array}{c}
  0  \\
  1   \end{array} \right)$,
 and (b) the time dependent adiabatic representation, which is given by the two instantaneous solutions $u(t)$ of
 $H(t,a)u(t)=w(t)u(t)$ with $w(t)$ the adiabatic energies. While the spin up/down representation has featured in many works
 (e.g., \cite {IrishGMS}-\cite {Ziegler}),
 the broader issue of representation choice in the density matrix has been intensively
 studied, e.g. in terms of the "einselection" in quantum measurements \cite{Zurek} and for the preference
  of energy states \cite{PazZ}.

 \end{enumerate}
  Further discussions of these steps are  given in the sequel.

\subsection{The significance of averaging in steps 1 and 2}
  Clearly, without an
 averaging the density matrices could be brought to a pure state form, with only one (diagonal) element
 unity and all the rest being zero.
 Illustrating this for an $N$ x $N$ density matrix when $N=2$, one can write
 the density matrix in the alternative forms
 \beq \left( \begin{array}{cc}
 a^* \cdot a &  a^* \cdot b \\
   b^* \cdot a  & b^* \cdot b
  \end{array} \right)\equiv \begin {array}{c}a^* \cdot \\ b^* \cdot \end {array} \left( \begin{array}{cc}
  a & b \\
  a & b
  \end{array} \right)\enq showing that the two rows are linearly
  related. Therefore one eigenvalue of the matrix is zero,
  while the other eigenvalue is, by invariance of the trace, \beq a^* \cdot a
  + b^* \cdot b =1 \label{1}\enq
  due to normalization. The density matrix can thus be brought to a
  form for a pure state. The same procedure holds also for any
  density matrix of size $N$ x $N$, with $N>2$, where the number
  of linear relations, and therefore of zero eigen-values is
  $N-1$. It is only when an averaging is performed {\it first} and
  the diagonalization of the averages is made subsequently, that a
  mixed state can arise and decoherence, with the vanishing of off-diagonal density
   matrix elements, emerges.

 However, to proceed literally as described in step 1, namely summing
 the density matrix over all (in practice, say, 1000)  initial phases of the oscillators
 would have meant to solve the TDSE 1000 times and to save all these solutions.
 Instead, as noted in step 2 above, we have solved it only once for one parameter set and
 averaged all density matrix elements over some time interval $2 \Delta t$. This time interval
 will be specified as we progress, the criterion being that an averaging over a greater interval
 does not alter the value of the averages. The relation
 of this procedure to an ensemble averaging is rooted in the
 ergodic theorem (or hypothesis) \cite{Farquhat}. We have also constantly checked our results for error
  and found that the full trace of the averaged
 density matrix deviated from (was short of) unity by less than $4\%$,
 though we have extended our computations over about 200 times the
 vibrational period ($2\pi/\omega$).
   Moreover, the normalization check of
 the wave propagated wave function was also in error by the same
margin (about $4\%$), indicating that the error in the density matrix has arisen
from numerical errors in the forward integration and not from
inadequate tracing (averaging).

\subsection {The "environment interaction"}
 A widespread formulation of the interaction of a bosonic
 environment with a spin system is to write the interaction of the spins with one
 or  more oscillators in the form  \beq \sum _n\sum_{i=x,y,z}  k^i_{n}q_{n}\sigma_i\label{intac1}\enq
 where $q_{n}$ is the $n$-th oscillator's amplitude and
 $k^i_{n}$ its coupling strength for the interaction with the spin \cite{LeggettEA, EnglmanY2004}.
 The behavior of the closed (spin-boson) system is studied
 through its density matrix $\rho_{s,b}$. The reduced density matrix of the
 spin system $\rho_s$ is then obtained from the trace $Tr_b\rho_{s,b}$ over all boson
  states and modes of the environment.

  What is the relation of this formulation to our model?

  In \er {H0} we have chosen  an (Einstein-) model for
 the oscillators, so that their frequencies are the same (denoted by
 $\omega$ and equated to $1$). However, the stochastic (random) effect of the
 environment on the spin systems is still present through each
 oscillator having a different phase $\alpha_a$, randomly distributed between $0$ and $2\pi$.
 We then replace the ${\it set}$ of randomly phased
 oscillators acting together by an ${\it ensemble}$ of independently acting
 oscillators, each oscillator having a phase $\alpha_a$, randomly distributed over the ensemble states.
 Here $a$ enumerates members of the ensemble.
 In summary, by the adopted semi-classical approximation for our model in \er{H0}, the oscillator amplitudes
 and coupling strengths in
 \er{intac1} take the (unnormalized) forms
 \beq
 q_a=\sin(\omega_a
 t+ \alpha_{a}),~k^x=k,k^y=k^z=0\label{class}\enq with $a$ labelling different states
 of the environment.
 In the von Neumann averaging in \ern{rho0}, it is the values of $\alpha_a$ that are to be
 summed over (eventually, integrated) .

 In the sense of spin-environment perturbation, the sine term  represents highly colored noise.

 In the next development of the present formalism,  noted in the previous subsection and in step 2 of section 2.1,
 the averaging over the solutions with differing
 initial phases  $\alpha_{a}$  has been replaced by averaging over a time interval $ 2\Delta t$.

\subsection{Decoherence Results}

\begin{figure}
\vspace{6cm} \includegraphics{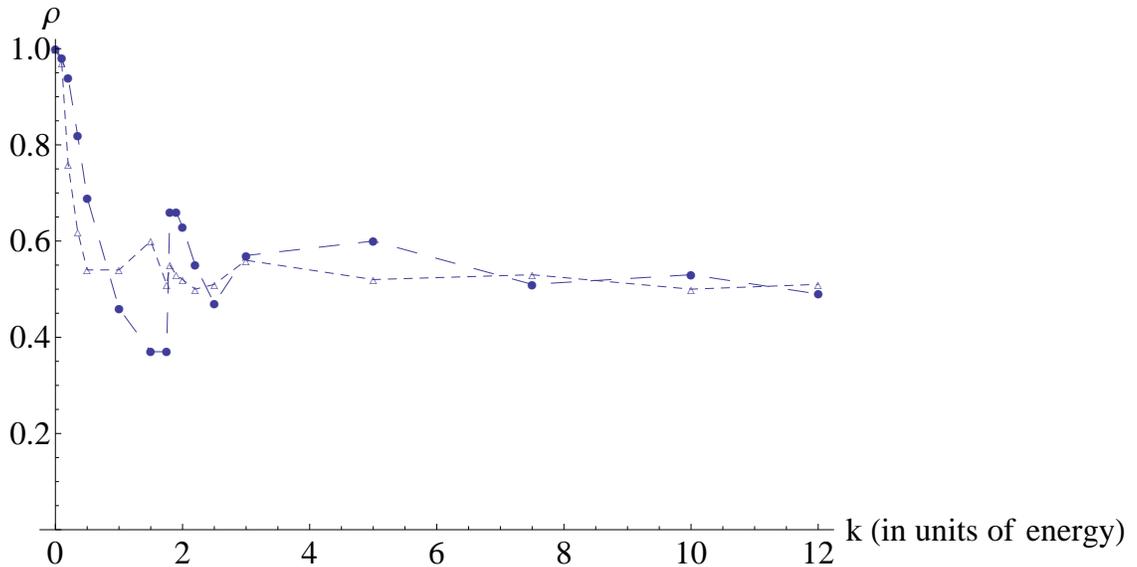} \caption {Decohered density matrix elements as
functions of the spin-classical oscillator coupling $k$ for small
spin energy $e=0.1$. Dots (connected for visual convenience by
long broken lines): Spin up probability. Triangles (connected by
short broken lines): Upper energy state occupation in the
adiabatic energy state representation.} \label{Decohfig1}
\end{figure}

The decohered diagonal matrix elements with time averaging over about five vibration periods and after reaching equilibration
 (such that longer times do not essentially change the average values and with near zero off-diagonal matrix elements, not shown) are presented
 in Figures 1-4,  (a) in the spin representation (broken line) and (b) in the adiabatic, time-instantaneous state
representation (dotted line). In figure \ref{Decohfig1} for weak
spin-oscillator coupling ($k<<1$) the initial (upper) state's
density matrix (mean occupation probability) is close to one, but
then decays to $\frac{1}{2}$ as the coupling $(k)$ increases. The
computed lower state's mean occupation probability was found to be
(1 {\it minus} the one shown), correct to about 0.001, verifying
the normalization of the density matrix. The limiting value of
$\frac{1}{2}$ is appropriate to equilibration with an oscillator
bath at infinite temperature, which pertains to this model.
(Finite temperatures and thermalization are treated in the next
section.).

There are sudden jumps, here as in the following figures, whose
nature is not clear, but probably reflect some resonances (i.e.,
occurrences when the instantaneous energy differences between the
states match the oscillator frequency, $\omega =1$).  To discount
the time-windows as the sources for the peaks (and also to provide
assurances for the reliability of the time averaging procedure),
we have consistently checked the accuracy of the averages, by
varying the time-window by 60-100 percents. The variations caused
changes in the time averages that were comparable to widths of the
line in our figures. Sharp variations in the state probabilities
were also seen for relatively small variations in the parameters
of the Hamiltonian in figure 9 of \cite{IrishGMS} (there termed
"unusual behavior").

\begin{figure}
\vspace{6cm} \includegraphics{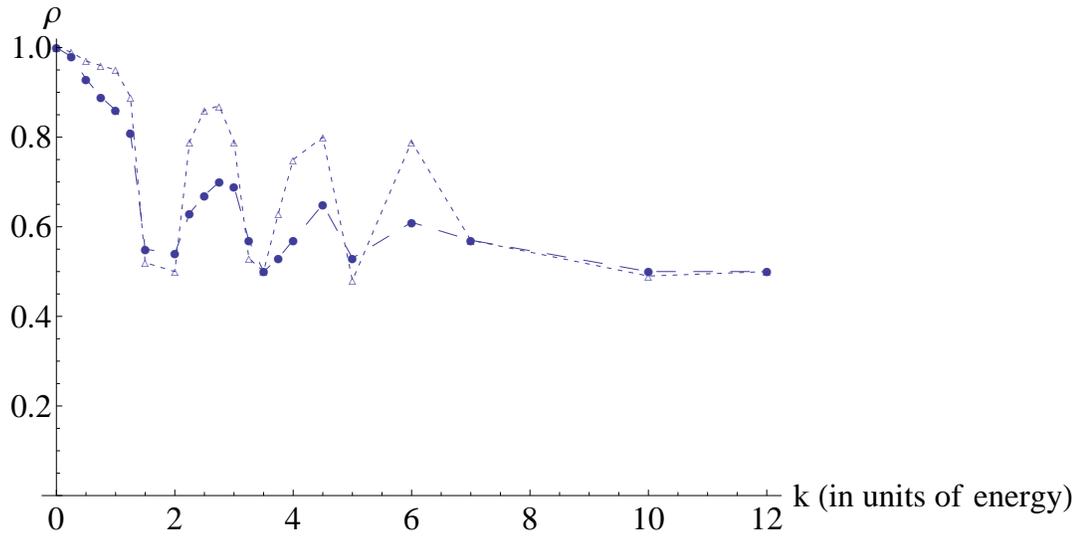} \caption {Decohered density matrix elements as
functions of coupling strength
 $k$ for moderate spin energy $e=1$. Dots: Spin up probability.
Triangles: Upper energy state occupation in the adiabatic energy
state representation.} \label{Decohfig2}
\end{figure}

In  figure \ref {Decohfig2} the same quantities are shown for spin energy $e=1$, of the same value as
the oscillation frequency. The equilibration starts for larger $k$ and the oscillations (resonances?)
are stronger.

\begin{figure}
\vspace{7cm} \includegraphics{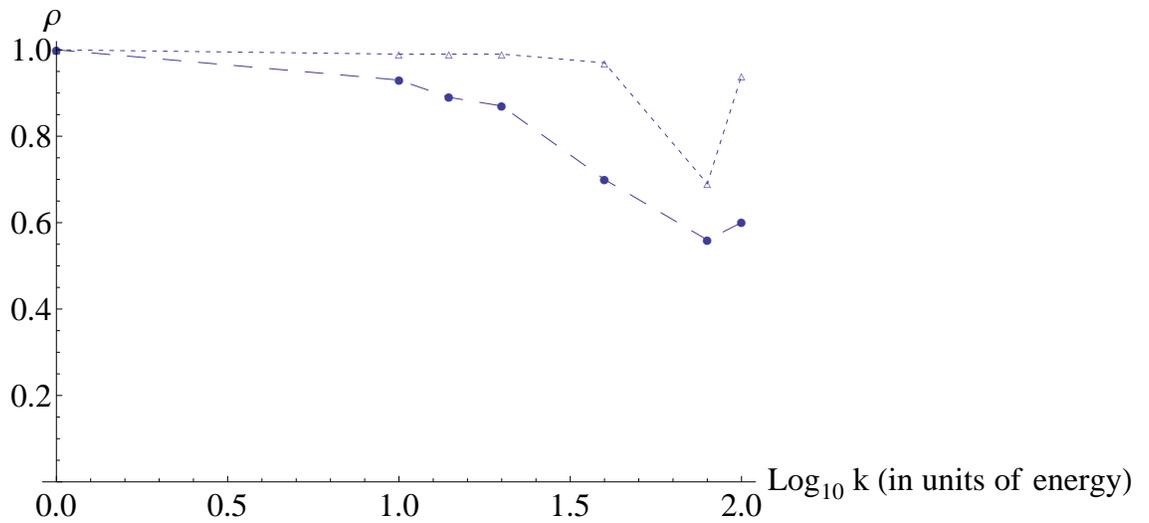} \caption {Decohered density matrix $\it vs$ $k$ in
the adiabatically protected spin (e=10) regime. Meaning of curves
as before. The coupling parameter $k$ reaches up in this figure
only to $2$, for much larger values the numerical results were not
reliable. } \label{Decohfig3}
\end{figure}

The third drawing, figure \ref {Decohfig3} is for spin energy $e=10>>1 =$ (the vibration frequency $\omega$), representing a situation, where the
adiabatic theorem holds, so  that there is no environment induced mixing of states. As seen, this holds
 for moderate values of the coupling, but for very strong coupling ($k>>1)$ the adiabaticity-protection breaks down.

\begin{figure}
\vspace{6cm} \includegraphics{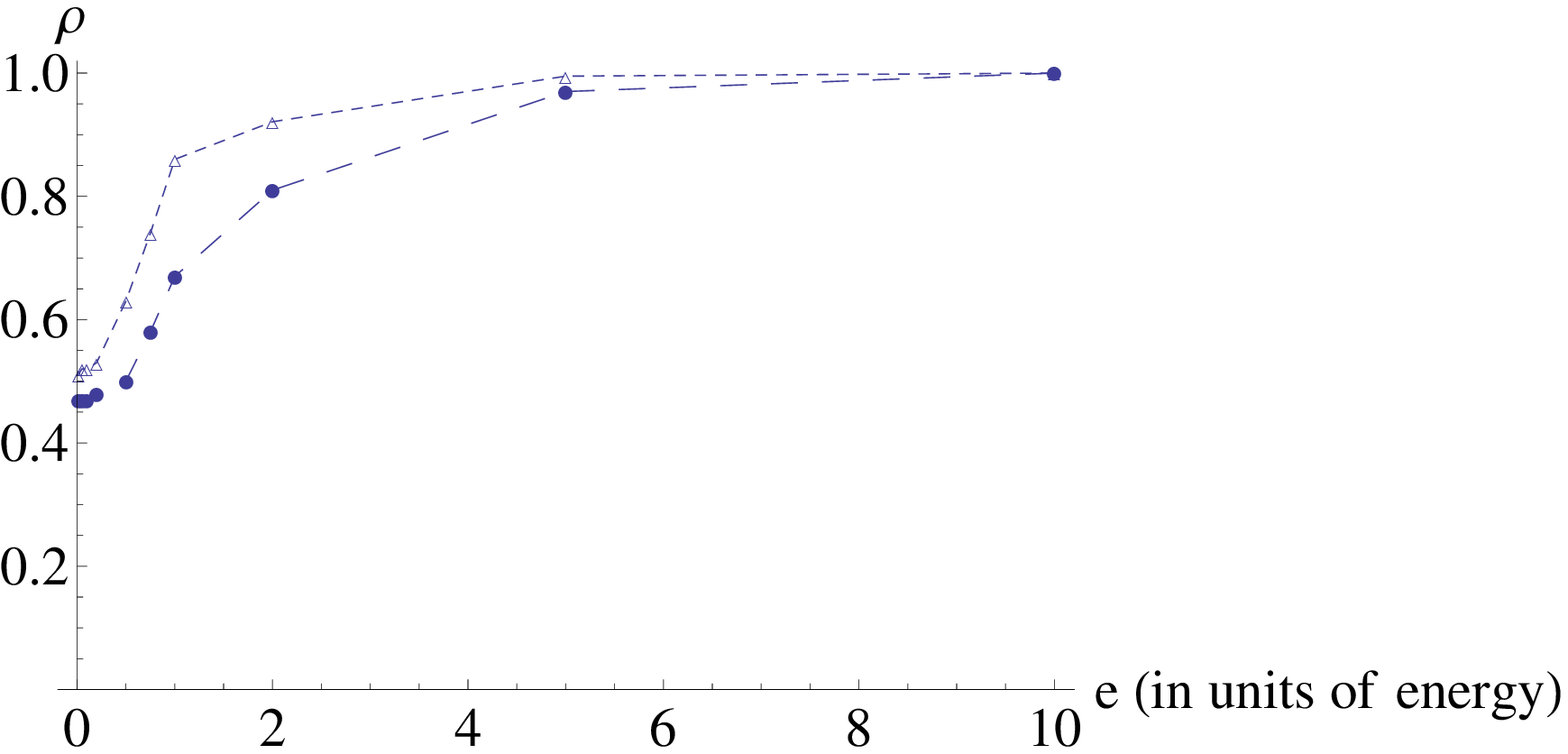} \caption {Decohered density matrix plotted against
spin energy $e$ for fixed coupling strength $(k=2.5)$. Curve
description, as before.} \label{Decohfig4}
\end{figure}

 Figure \ref {Decohfig4} shows the inverse situation that the coupling strength is held fixed at $k=2.5$ and the
 spin energy is varied from $e \simeq 0$ (equilibrated case) to a large value (the adiabatically protected regime).

\subsection {Thermalization}

While the former results  in figures 1-4 showed decoherence at
essentially infinite temperature ($T \equiv 1/(k_B\beta)~ ->\infty
$), or with the same probability for up and down flipping by the
oscillator, at finite temperatures the two probabilities differ.
Since in our model stochasticity enters through time-averaging, we
include temperature effects by weighting the time duration
according to the energy of the system, meaning that excursions at
lower energies have greater time-weights than those at higher
energies. Quantitatively, we express \er{rho2} in the adiabatic
energy representation, in which $m(t)$ indexes the two states in
the adiabatic representation. Then replace the diagonal terms in
\er{rho2} by: \beq \rho_{mm}(t) = \frac{1}{2\Delta
t}\int_{t-\Delta t}^{t+\Delta t}d\tau \frac{e^{-\beta
E_m(\tau)}|<m(\tau) |\psi(\tau)>|^2}{<\psi(\tau)|e^{-\beta
H(\tau)}|\psi(\tau)>}
 \label{rho3}
 \enq
 in which the denominator in the integrand ensures the normalization of the density matrix.
 The off-diagonal elements are negligible also in the thermalized density matrix.

 An intuitive justification for the chosen time-weighting can be
 based on the early work Rechtman and Penrose \cite{RechtmanP}, who
 have shown that the probability distribution for a finite
 classical system, in thermal contact with an infinite (in practice,
  sufficiently large)
  heat bath, with the composite  system being distributed
  micro-canonically, is the Gibbs canonical distribution $e^{-\beta
  E}$, where $E$ is the energy of the system. This has the meaning that for
  a state of energy $E$ of the system and no degeneracies, the
  number of micro-states of the heat-bath is proportional to $e^{-\beta
  E}$. If we now suppose that the heat bath spends equal {\it time} in
  each micro-state (cf. the ergodic hypothesis), then in the system's time integration
  the infinitesimal $dt$ has to be weighted by the Gibbs factor,
  as in \er{rho3}.

\begin{figure}
\vspace{6cm} \includegraphics{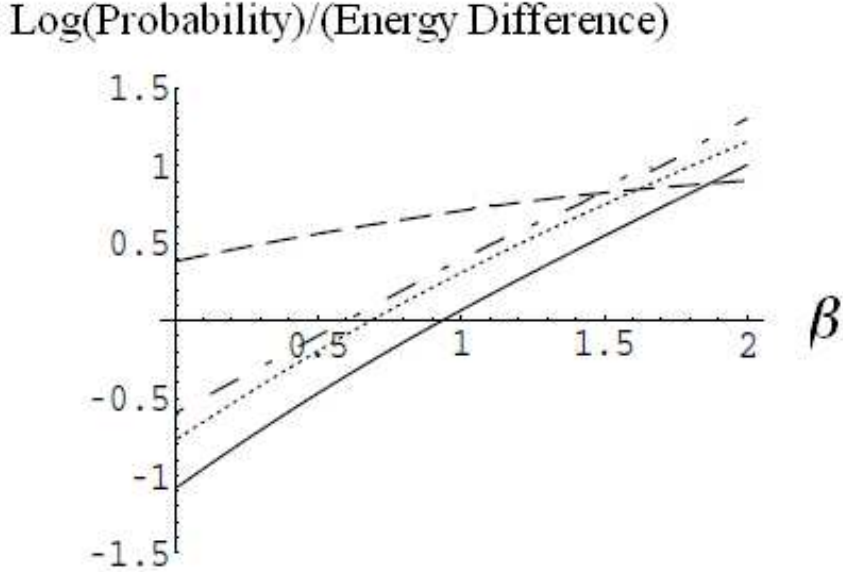} \caption {Thermalized density of states. The
input parameters for the curves from bottom to top (at $\beta = 0$
or infinite temperature) are the following: Solid line: spin
energy  $e = 0.5$, coupling $k = 1$, time*frequency = 270,
vertical displacement = 0, (in this curve the mean slope is 1).
Dotted line: $e = 0.5$, $k = 1$, time*fr. = 60, vertical
displacement = 0.2. Chained line: $e = 5$ (near adiabatic), $k =
0.2$, time*fr. = 250, vertical displacement = 0.6. Long broken
line: $e = 0.1$, $k = 2.5$, time*fr. = 250, vertical displacement
= 0.4. Curves are displaced vertically for  visual clarity.}
 \label{Decohfig5}
 \end{figure}

 To check the validity of our proposed thermalization procedure, we investigate whether
 we regain through it the "$\rho_{mm}=e^{-\frac{E_m}{k_B T}}/Z$" law?  In figure \ref{Decohfig5} we
 show the logarithm of the ratio  of the up and down (time-averaged) diagonal density
 matrices in the adiabatic  state representation, divided by the
adiabatic energy difference, against the inverse temperature $\beta \equiv \frac{1}{k_B T}$.
 In thermalized energy eigenstates the plot should be linear in $\beta$ with a slope of one. This is
 approximately the case for
the three lower curves (in which $k\leq1$, weak to moderate spin-oscillator coupling), but for the
uppermost curve (in which $k=2.5$) the spin is too much interwoven with the environment to thermalize
independently of it.

\subsection{Revivals}
\begin{figure} \vspace{6cm} \includegraphics{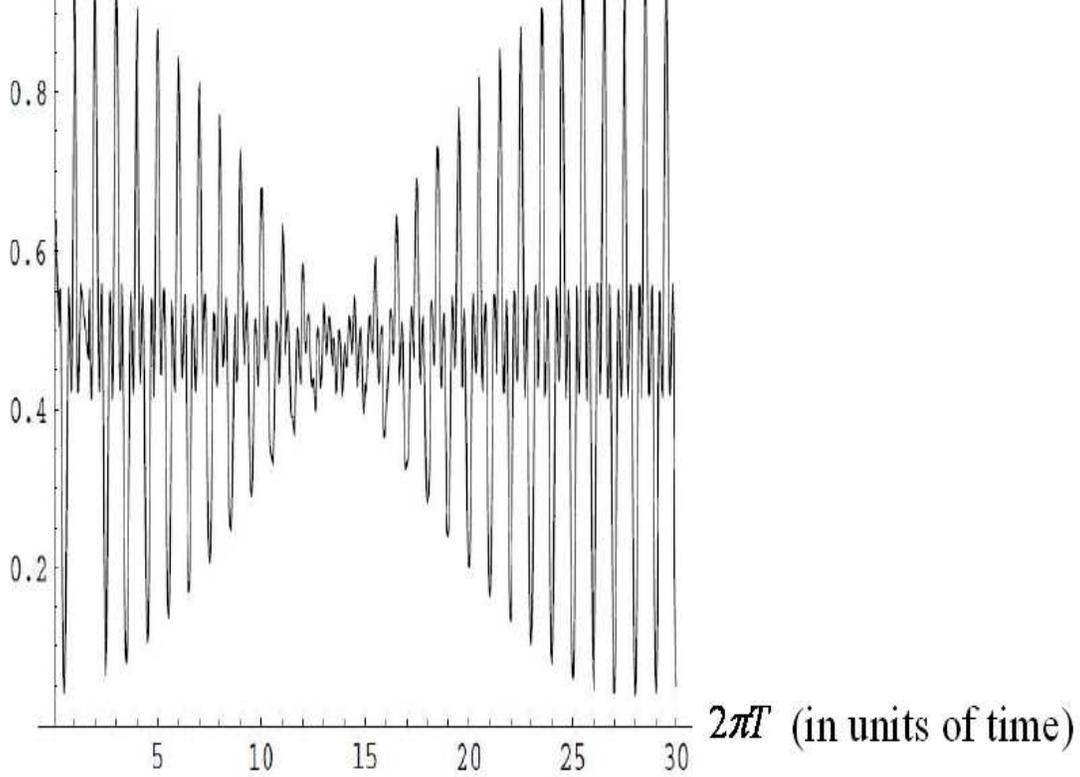} \caption {Time-averaged spin-up probability. Model parameters: $e$, spin energy
$=.05$, $k$ (the spin-oscillator coupling strength) $= 2.5$, time averaging over $1.2/\omega$, $\omega =1$.}
\label{revival}
 \end{figure}

To establish the compatibility of our approach with previous works
(some of them analytic) on the Rabi model, we turn to a study in
which the oscillator state was modelled by a coherent state (\cite
{IrishGMS}, section III). As is well known, coherent quantum
states resemble closely the behavior of a classical oscillator
(which features in our Hamiltonian). Long period (compared to the
oscillator's period) revivals in
 the up (or down) spin-state probabilities (equivalent to the diagonal terms in the reduced density
 matrix) were shown in Figure 7(a) of that paper in the adiabatic limit. The curve computed by us and
  shown in Figure \ref {revival} (persistent for many further periods, not shown)
 is extremely similar to their result for a coherent state.
Our parameter choice ($e$, spin energy $=.05$, $k$,
spin-oscillator coupling strength $= 2.5$) is not immediately
translatable to that ($<N>=1, \lambda/\omega =0.1$) in \cite
{IrishGMS}, since the coupling strength $k$ in our semi-classical
formalism is a combination of these. We have also found that the
complete revival pattern shown in Figure \ref {revival} occurs for
only a restricted choice of parameters and is not a universal
feature of the model. However, referring to the drawings (a) to
(e) in figure 9 of
 \cite {IrishGMS}, we note that also in their model even slight changes of the
 parameters cause radical changes in the patterns.

\section{Two Qubit Systems}
The following Hamiltonian $\textbf{H}_{total}(t)$ involves two
half-spin systems (qubits), whose parameters are in the sequel
consistently designated by capital and lower-case letters,
respectively, interacting with a classical boson source (as
before, vibrational or light-like) varying with time. \ber
 \textbf{H}_{total}(t) & = &  H(t) +h(t) +
\textbf{H}_{int} \label {Htotal}\\
H(t) & = &  E\Sigma_z + K\Sigma_x \sin(\omega t+\alpha)\label{H}\\
h(t) & = &  e\sigma_z + k \sigma_x\sin(\omega t+\alpha') \label{h}\\
\textbf{H}_{int} & = & \gamma (\Sigma_z \cdot
\sigma_z+\Sigma_x\cdot \sigma_x) +\gamma' (\Sigma_x \cdot
\sigma_x+\Sigma_y\cdot \sigma_y)\label{Hint}\enr having written in
the first line the total Hamiltonian, comprising, as in the
following lines, the Hamiltonians of the large- and the
small-symbol system, ending with the interaction between these.
 In them $E,e$ are energies of the two spin systems, the $\Sigma$'s (!) and $\sigma$'s are Pauli matrices operating in the
respective spin-spaces, $K,k,\alpha$, are parameters of the spin-boson couplings and $\omega$
is the frequency of the interacting source. The external boson source is classical and for
it the Hamiltonian need not be written out.

The strengths of the Spin-spin interaction are denoted by $\gamma$ and $\gamma'$.
 For the form of the interaction two alternative sub-models will be used:
 The first, named "The two-dimensional model", for which $\gamma\neq 0$ and $\gamma'=0$,
is fashioned after the $E\bigotimes e$ vibronic interaction
\cite{Ebook}. The second, in which $\gamma'\neq 0$ and $\gamma
=0$, commonly features in Ising models and is known as the
"transverse interaction".
 It has been recently used for superconductors with a large pseudo-gap and weak long rage Coulomb
 interaction \cite {Feigelman}.

The double inequality (exemplified with a physical model in the
Introduction):
 \beq
 E>>\hbar \omega >>e
 \label{inequ}
 \enq
is the keynote to the present section, in that it makes the time
variation in the Hamiltonian slow (adiabatic) with respect to that
of one of the spins (the "capitalized" one) and fast
(non-adiabatic) with respect to that of the other  ("the
lower-case" one). It is therefore expected that the distilled wave
function in the former's Hilbert space will stay coherent, while
that one in the latter's Hilbert space will decohere. This result
is indeed found, with some interesting features to be expatiated
on in the sequel. (Two spin systems with {\it identical} energies
were treated in the adiabatic limit in \cite {YangZZ}.)

We wish to investigate decoherence in the combined system.
 In an overwhelmingly large number of papers "Decoherence", leading from an initially pure to a
later mixed state, has been  obtained by going from the density of
states in the full to a partial Hilbert space, through tracing
over the complementary Hilbert space (e.g, references in
\cite{AllahverdyanBN}). As noted earlier in the programmatic
summary of section 2.1, we use an alternative procedure for
decoherence, namely  an external parameter averaging procedure. In
terms of our model, in which the environment is represented by a
classical oscillator, this means that after obtaining a
(time-dependent) solution $\psi_{\alpha,\alpha'}(t)$ for  given
phases $\alpha,\alpha'$, we average the elements of the density
matrix $\rho_{nm}(t)$, with respect to all values of these
parameters. Then, formallly \beq
\rho_{nm}(t)=(4\pi^2)^{-1}\int_0^{2\pi}d\alpha\int_0^{2\pi}d\alpha'<n|\psi_{\alpha,\alpha'}(t)>
<\psi_{\alpha,\alpha'}(t)|m>\label{rho1}\enq for some chosen
representation, whose components are labelled $(n,m)$. This
procedure differs from the commonly used ones (e.g., the
stochastic Schr\"o\-dinger equation or a time-propagation equation
for the reduced density matrix) in which the environment is also
in a quantum state, whose nature is specified by its spectral
properties \cite {BieleD}. Still the method used here is
historically primordial (coming from \cite{Neumann, Band}). It is
also suitable for numerical calculations and alleviates to some
extent the classical-quantal dichotomy, extensively treated in
\cite{AllahverdyanBN}.

The four component wave function $\psi$ is inserted from the
numerical solution of the time dependent \SE ~(with $\hbar=1$)\beq
i\frac{\partial \psi_{\alpha, \alpha'}(t)}{\partial t}=
\textbf{H}_{total}(t)\psi_{\alpha, \alpha'}(t)\label{TDSE}\enq
 with the appropriate initial condition at $t=0$. This system is then in a pure state; to track
down its progression in interaction with a stochastic environment
towards a (possibly) mixed state, we need to consider the density
matrix of the system.

As already remarked, the density matrix is
representation dependent (though its trace is not), and the choice
of the representation (labelled above $nm$) for the density matrix
 was extensively discussed in several
publications, e.g. \cite {Zurek}. The conclusion there was that an
"environment induced selection (einselection)" takes place due to
the (experimentalist's) choice of the pointer, which is expressed
by the form of the interaction between the environment and the
system. In this choice, the off-diagonal matrix elements of the
system's density matrix vanish in a time shorter than other time
scales in the system's Hamiltonian. (It will be seen that our
numerical results support their choice for "einselection".)
Furthermore, under conditions of weak coupling and large energy
scales it was formally shown in \cite {PazZ} that the choice
pointer states are the discrete energy states of the system. In
this context, one recalls an early, somewhat enigmatic statement
in \cite {DanenLP}: "In general, only quantities quasi-diagonal in
the energy representation are observable".

 This has dictated the choice for one of our two adopted representations
 ("b" in the programmatic summary in section 2.1) as the adiabatic solutions of
the Hamiltonian, namely the instantaneous (upper and lower energy)
solutions $u(t),l(t)$ for the small-energy part of the Hamiltonian
\cite{Messiah}, i.e., \beq h(t)[u(t)/l(t)]=w_{u/l}(t)
[u(t)/l(t)]\label{hadiabat}\enq  and likewise the adiabatic
solutions $U(t)$,$L(t)$ for the large- energy part of the
Hamiltonian: \beq H(t)[U(t)/L(t)]=W_{U/L}(t)
[U(t)/L(t)]\label{Hadiabat}\enq Thus the $4$ x $4$ density matrix
is written in the representation of $Uu,Ul,Lu,Ll$ in the given
order.The appropriate initial condition is an energy eigenstate at
$t=0$. The alternative choice for the representation, namely  the
more conventional spin up/down representation ("a" in section
2.1), is not treated in this, two-qubit section, since we could
not find results in the literature to which we might make
comparison..

 Actually (as already indicated in section 2.1), for the sake of simplifications
  in our procedure to obtain the density  matrix at any time $t$,
 we have averaged not over the initial parameters $\alpha,
 \alpha'$, but rather, with fixed values of these
 $\alpha=0=\alpha'$, over a spread of the times ($t-\Delta t,t+\Delta
 t$), $\Delta t$ being in the two-qubit case close to the oscillator period-squared
 $(2\pi/\omega)^2$, or about $40$ in our time units ($\omega=1$) (see integral in \ern{rho1}).

\subsection{Non-interacting spin systems}

\begin{figure} \vspace{6cm} \includegraphics{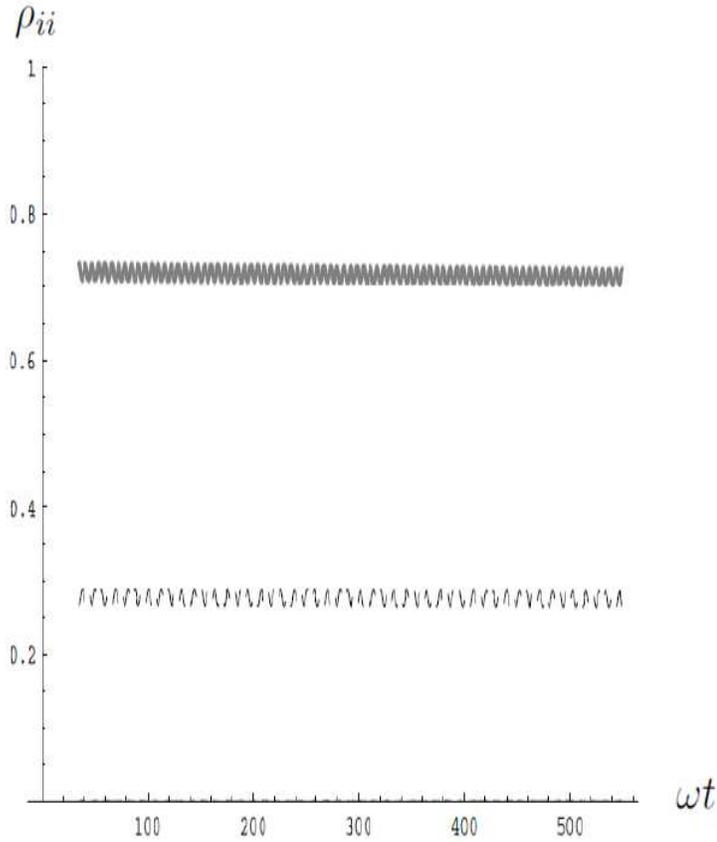} \caption {Diagonal density matrix
elements $\rho_{ii}~ ( i=[Uu,Ul,Lu,Ll])$ {\it vs } normalized time.
Time averages of overlap squares are shown in the adiabatic
representation described in the text. Only the diagonal matrix
elements for $i=Uu,Ul$ (shown in this order from above to below) are visible, while those for $i=Lu,Ll$ are
too small to be visible without magnification. Since at any reduced time $\omega t$
the averaging spreads over $\pm 30$, the first shown value is above
30, thus missing the starting values for the four component
wave-function (1,0,0,0). The constancy of the averages and the small
deviations from these throughout the time range are to be noted.}
\label{PDfig1}
 \end{figure}

As a start, we consider the simplified situation in which the spin
systems do not interact, i.e., that $\gamma =\gamma'=0$ in
\ern{Hint}. Although this case can be treated for the two spins
separately, for the sake of continuity with the interacting spin
case in later sections, we treat the two spins as belonging to a
larger, combined Hilbert space. We show below the resulting $4$ x
$4$ density matrix obtained, as described above, from averaging
over neighboring times (by an integration over about 10 oscillator
periods)
 and then further representing the obtained
averages by their mean values over the full computed time range
(in practice: about 500 vibrational periods), together with
specification of the standard deviation of the values inside this
time range.  Obtained results are shown in Figure \ref{PDfig1} for
the chosen parameter values of
 \beq E=5,e=0.1,K=2,k=0.125,
\alpha=\alpha'=0,\gamma=\gamma'=0 \label{param} \enq in units of
$\omega$ and they are characteristic of other parameter values.

The time averaged density matrix is shown next:
  \beq <\rho>= \begin{array}{c}
  <Uu|:\\<Ul|:\\<Lu|:\\<Ll|:
    \end{array} \left( \begin{array}{cccc}
  0.721 \pm .015 & 10^{-2} & 10^{-4} & 10^{-4} \\
  10^{-2}& 0.272 \pm .010 & 10^{-5} & 10^{-3}\\ 10^{-4} & 10^{-5}  & 2\cdot10^{-4} & 10^{-6}\\10^{-4} & 10^{-3}&10^{-6} &
  4\cdot 10^{-4}
  \end{array} \right)\label{rho1b}\enq
  The $\pm$ deviations represent estimated variations in the values over
  the whole time investigation range. Their signature in Figure \ref{PDfig1} are the small wiggles on the
  otherwise horizontal lines. The off-diagonal entries
  show upper limits to absolute values. The deviations are partly due to
  computational inaccuracies, partly to the finite range of the averaging process
    and partly to the parameters not being in the extreme adiabatic limit.

   [It may be added that the above error-checking refers exclusively
to the diagonal terms in the density matrix,  whereas the {\it
averaged} off-diagonal matrix elements were unexceptionally
negligible. This means that the (adiabatic, instantaneous)
representation used here was indeed the proper ("einselected")
one. These results then lend  numerical support for the analytical
arguments of  Paz and Zurek\cite {PazZ}.]

  \subsubsection{Reduced density matrices}

  With the complementary subsystem traced over, the reduced density matrices
  for the small and large energy systems are (with suppression of errors), respectively :

\beq <\rho_{u/l}>\approx \begin{array}{c}
  <u|:\\<l|:
    \end{array} \left( \begin{array}{cc}
  0.721 &  0 \\
   0 & 0.272
  \end{array} \right)\label{rhoul}\enq

\beq <\rho_{U/L}>\approx \begin{array}{c}
  <U|:\\<L|:
    \end{array} \left( \begin{array}{cc}
 0.993 &  0 \\
   0 & 0
  \end{array} \right)\label{rhoUL}\enq

The small energy system is thus seen to have decohered, or be in a
mixture state (with the  partitioning of the weights depending on
the values of the parameters, $e,\omega, k$); while the large
energy state is throughout in a coherent, pure state, due to its
protectedness by the adiabatic theorem. For situations not
belonging to the extreme adiabatic limit (represented  by
$\frac{E}{\hbar\omega} \longrightarrow \infty $), there will be a
finite decoherence time, in the course of which the pure state
also decoheres (equilibrates). This decoherence time will decrease
as the above ratio decreases, but in our computation range
(typically 500 vibrational periods) we have not found for the
high-energy (adiabatically protected) states a finite decoherence
time.

When the environment coupling to the small energy system was enlarged to $k=0.5$
(instead of $k=0.125$, in the previous case), the diagonal matrix elements took the {\it mean}
values, with their standard deviation not noted, $[0.804,0.192, .0004, .0000]$

\subsection{Systems with Spin-spin interaction}
\subsubsection {Two-dimensional sub-model,$\gamma\neq 0,\gamma' = 0$}
{\it ~ Weak interaction:}

\begin{figure} \vspace{6cm} \includegraphics{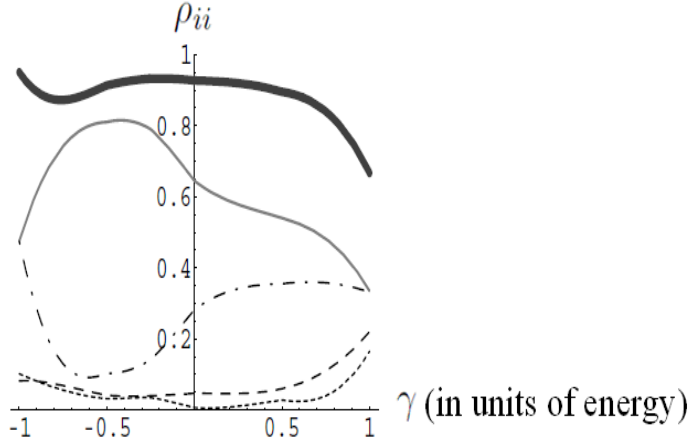} \caption {Diagonal density matrix
elements $\rho_{ii}, i=[Uu,Ul,Lu,Ll]$ (for the thin curves in this
order from top to bottom)
 as function of a weak $\gamma$ (the 2D Spin-spin coupling strength). The strongly drawn curve
 is the sum of $Uu$ and $Ul$, giving the diagonal U term in the reduced density matrix.}
\label{PDfig3S}
 \end{figure}
\begin{figure} \vspace{7cm} \includegraphics{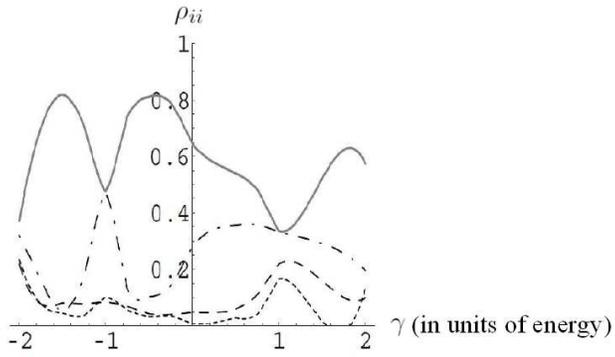} \caption {Diagonal density matrix
elements against  $\gamma$, the 2D Spin-spin coupling strength for
extended values of the coupling. Meaning of curves as in previous
figure.} \label{PDFigure3}
 \end{figure}

We first investigate how does the interaction between the spin
systems modify
 the decoherence discrimination between adiabatic and non-adiabatic systems.  In Figure
\ref{PDfig3S} the diagonal density matrix elements are shown for
$|\gamma|\leq 1$.
 Let us set, somewhat arbitrarily the criterion for decoherence discrimination between
 large and small energy states as a $80\%$ purity for the large energy states. The thick line
 in Figure \ref{PDfig3S} shows the
 sum of the two uppermost thin curves (for $Uu$ and $Ul$) in that figure: One sees that
  the $>0.8$ criteria for purity is well satisfied for negative couplings in the range
 $0>\gamma>-1$, but does not hold near the upper values in the positive range $0<\gamma<1$, for which
 the two curves do not add up to $0.8$.
  One also notices that for $|\gamma|\approx 1$ the small energy states ($u$ and $l$) are "fully" mixed,
  i.e. their diagonal values are equal. However, this does not hold for higher coupling strength, as
  Figure \ref{PDFigure3} illustrates.

~ \vfill \eject \noindent {\it Higher interaction strengths:}
\\ \\
 In Figure \ref{PDFigure3} one sees that as the coupling strength $\gamma$ is varied
 a high level of weight exchange takes place between the terms, especially between the
 diagonal $Uu$ and $Lu$ terms. One notes signs of the "level crossing avoidance" phenomenon, familiar
 from energy level plots for interacting states.

\subsection {Ising coupling model, $\gamma= 0,\gamma'\neq 0$}
\begin{figure} \vspace{6cm} \includegraphics{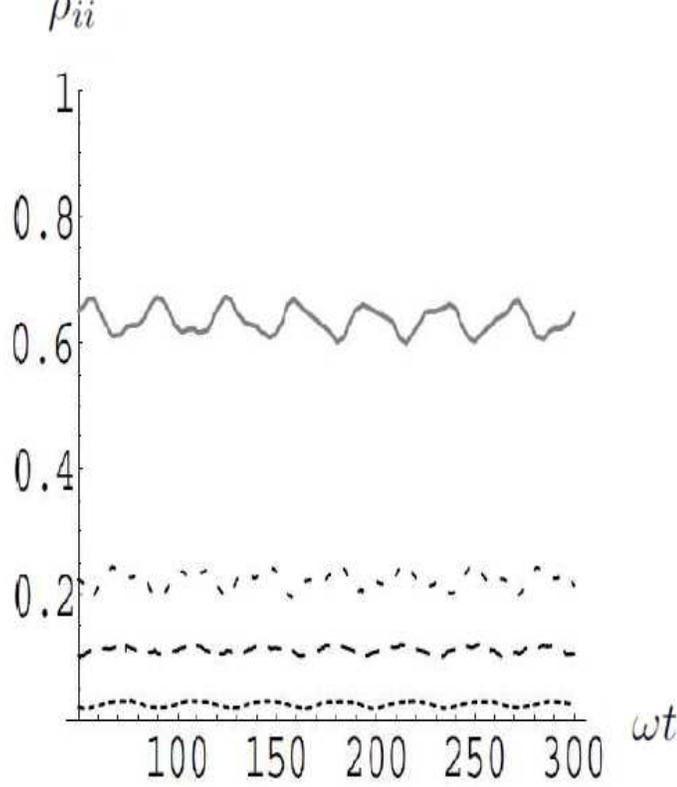} \caption {Time-asymptotic diagonal density matrix
elements under moderate Spin-spin interaction ($\gamma'=1$) in the Ising model coupling plotted against normalized time.
Results similar to those in Figure \ref {PDfig1}, but with more "noise". However, there is an {\it overall}
 decoherence as a  result of the
coupling. The diagonal matrix elements are from top to bottom for
$Uu,Lu,Ul,Ll$. For the parameter set:[$E=5, e=0.1, K=4, k=2.5,
\omega=1,\gamma=0, \gamma'=-0.5$]} \label{PDFigure4}
 \end{figure}

Numerical results are shown in Figure \ref{PDFigure4} (for parameters $E/\omega=5,e/\omega=0.1,K/\omega=4,k/\omega=2.5)$

 \subsubsection {Remarkable appearance of "revivals"}
\begin{figure} \vspace{6cm} \includegraphics{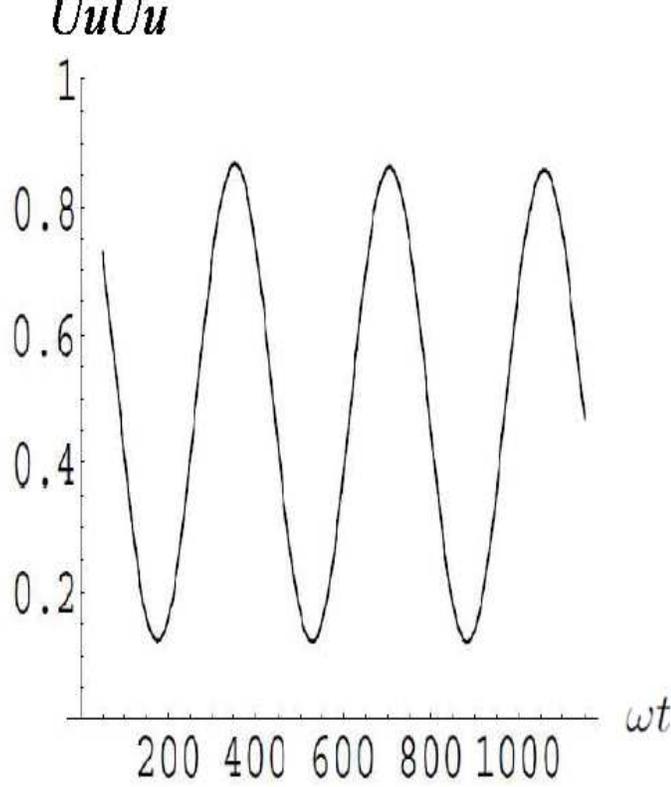} \caption {Large scale, large period ($>>2\pi/\omega$) oscillations in the diagonal
 Uu-component of the density matrix, superimposed on tiny $\approx 2\pi/\omega$ oscillations (not visible).
Parameter values:[$E=5, e=0.1, K=4, k=2.5, \omega=1, \gamma=0,
\gamma'=-0.5$]} \label{PDoscil}
 \end{figure}

 "Revivals" or large amplitude - long period returns to the starting diagonal elements in the density of states
 have been shown in section  2.6 for a single qubit, Rabi model. Similar phenomena occur also in the two qubit case.
 While for most values of the parameters the diagonal density matrix elements exhibit only small
 oscillations over the time range in the asymptotic, long-time range, typically $\delta\rho \approx1\%$,
 there are some singular values of the parameters where the oscillations are of the order of $100\%$ and persist
 over several periods.The periods are in the range of $50\cdot 2\pi/\omega$ or larger. An example of
  this behavior is shown in Figure \ref{PDoscil} for the parameter set [$E=5,e=0.1,K=4,\omega=1, k=2.5$]
   at coupling strengths in the close vicinity of $\gamma'=-0.5 $, but not at more than about $0.02$
   away from this value.  Similar oscillations with comparable periods are observed for the parameter set
   [$ E=5, e =0.1, K=2, k =1.25, \omega=1$], near the coupling strength value of $\gamma'=.8$.

   The oscillations are the more remarkable in that the time period of $300/\omega$, does not have any
   simple physical explanation in terms of the parameter set. (This is unlike the revival time
   expression in \cite {IrishGMS}, holding for weak coupling and large $<N>$ case). Further investigation
   is needed to reveal the source of this result.

   \section {Discussion}

    In a two-qubit system whose energy-splittings are (respectively) much larger and much
    smaller than the frequencies in  externally induced time-dependent perturbations, the low energy qubit
    decoheres, while the high energy qubit maintains its purity, being adiabatically protected. While this
     may be intuitively obvious, we have also examined less obvious cases when the two qubits are coupled
    and have found that for large coupling strength the adiabatic protection wears off.

    The time averaging procedure used in the paper (introduced programatically in section 2.1)
    has been found to be in practice more time-economic for achieving the density matrix
     truncation than the mainstream formalism of environment
     tracing. A more basic advantage is that whereas in the latter
     method the truncation has to be enacted by some assumption of
     the nature and dynamics of the environment (e.g., the
     measuring apparatus, as in section 1.1.2 in \cite
     {AllahverdyanBN}), which is outside and beyond the
     Hamiltonian of the system itself, in the present formalism
     the same effect is achieved by the relatively simple
     manipulation (time averaging) of the {\it system's} Hamiltonians, as those
     in \er{H0} and \er{Htotal}-\er{Hint}, without speculating on
     the time development of the environment. In this
     sense, then, the system's Hamiltonian is much more
     self-contained than that of the mainstream formalism.
     The choice of the time dependent terms in the above equations is
     apparently {\it ad hoc} and arbitrary; we have also tried in
     a preliminary way other forms (like a sum of oscillator terms, with
     random amplitudes and frequencies) and the results appear to
     be similar, except for the required time-width for averaging.
     To present these results in a systematic form represents an
     extension of the present (single frequency) model, which was
     indeed named "A Minimal Model" in a recent presentation \cite{EnglmanY2012}.

Another aspect for which the present formalism fills a gap is the
distinction for the density matrix truncation, also discussed in
section 1.1.2 of \cite {AllahverdyanBN}, between single run
experiments (which are the ones frequently made in the laboratory
and which are directly addressed  by our formalism) and multiple
run experiments (formalized by ensemble tracing). It is beyond the
scope of the present work to investigate how time averaging in a
single run is to be carried out in practice.

A further evident ramification of the time averaging formalism is
in the direction of the second moments of the density matrix. (At
present only the averages, or first moments, feature in our
results.) This would be of importance, e.g.,  for expressing the
"basin of equilibration" (quantified in Eq. 8 of \cite{LindenPSW}
through the ratio of the "effective dimension explored" by the
environment to the system's environment), in terms of the
parameters of our formalism.

\end{document}